\documentclass[a4paper]{jpconf}
\usepackage{graphicx}
\usepackage{amsmath}
\begin{document}
\title{Search for exotic states in photoproduction at GlueX}

\author{Abdennacer Hamdi (on behalf of the GlueX Collaboration)}

\address{GSI Helmholtzzentrum f\"ur Schwerionenforschung GmbH, Planckstr. 1, 64291 Darmstadt, Germany}

\ead{a.hamdi@gsi.de}

\begin{abstract}
    Quantum Chromodynamics (QCD) is the theory that describes how hadrons are built from quarks and gluons via the strong interaction. Many predictions have been experimentally confirmed, but others remain under experimental investigation. Of particular interest is how gluonic excitations give rise to states with constituent glue. One class of such states are hybrid mesons that are predicted by theoretical models and Lattice QCD calculations. Searching for and understanding the nature of these states is a primary physics goal of the GlueX experiment at the CEBAF accelerator at Jefferson Lab in the US. We will give an overview of the experiment, and present the status of the search for a hybrid meson candidate, $Y(2175)$. This work is supported by HGS-HIRe.
\end{abstract}

\section{Introduction}
Mesons in the constituent quark model are color-singlet bound states of a quark $q$ and antiquark $\overline{q}$, with quantum numbers $J^{PC} = 0^{-+}, 0^{++}, 1^{--}, 1^{+-}, 1^{++}, 2^{--}, 2^{-+}, 2^{++}, etc$, where $J$, $P$ and $C$ are total angular momentum, parity and charge conjugation of the fermion system, respectively. This simple picture has successfully described many observed states in the meson spectrum. However a richer spectrum is allowed by Quantum Chromodynamics (QCD) that includes the gluonic degrees of freedom in the quark and anti-quark system. Since the gluonic field can carry different quantum numbers this introduces many new states to the spectrum, including those carrying quantum numbers: $J^{PC} = 0^{--}, 0^{+-}, 1^{-+}, 2^{+-}, etc$, that are not allowed for conventional $q\overline{q}$ mesons. These latter are the (spin-)exotic hybrid mesons, and their experimental observation will be a proof of the existence of such states beyond the constituent quark model. The hybrid mesons are predicted by many phenomenological models~\cite{ref.1}, and Lattice QCD is making predictions for their properties like the mass~\cite{ref.2}, which can be tested experimentally.

\section{Experimental Setup}
\begin{figure}[h]
    \centering
    \includegraphics[width=25pc]{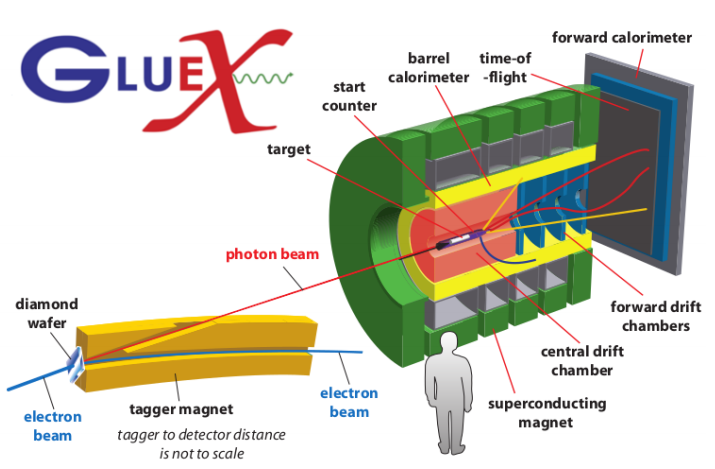}
    \caption{\label{fig.1}The GlueX experimental setup.}
\end{figure}

The GlueX experiment is dedicated to the mapping of the spectrum of hybrid mesons, using a high-energy linearly polarized photon beam produced by a 12 GeV electron-beam through coherent bremsstrahlung on a diamond radiator. By choosing the crystal axis orientation of the diamond we produced four data sets, with 2 sets of parallel (PARA) ($0^{\circ}/90^{\circ}$) and perpendicular (PERP) ($45^{\circ}/135^{\circ}$) polarization orientations, respectively. The energy and intensity of the photon beam are monitored by a pair spectrometer system (dipole magnet and scintillator arrays), and to measure the polarization, a triplet polarimeter ($\gamma e^{-} \rightarrow e^{+}e^{-}e^{+}$ scattering on a thin Be foil) is used. The photon beam impinges on a 30 cm long liquid hydrogen target positioned along the central axis of the detector (see Figure~\ref{fig.1}). The central region of the detector is contained in a solenoid magnet with $\sim$ 2T on its central axis. Particles from the primary interaction first pass through the Start Counter (scintillator detector), which helps identify the beam bucket which generated the event. Directly surrounding the Start Counter is the Central Drift Chamber (CDC) (straw tube detector), providing tracking and energy loss ($dE/dx$) information. Downstream of the CDC are the four packages of the Forward Drift Chamber system (FDC) (planar drift chambers), providing tracking as well as the $dE/dx$ information. Surrounding the tracking devices is the Barrel Calorimeter (BCAL) (lead scintillator fiber) sensitive to photons between polar angles of $11^{\circ}$ and $126^{\circ}$. Downstream of the solenoid is the Forward Calorimeter (FCAL) (lead-glass blocks), which covers polar angles from $1^{\circ}$ to $11^{\circ}$. In front of the FCAL is the Time Of Flight wall (TOF) (scintillator bars) providing timing information. For more details about the GlueX detector components and performance see~\cite{ref.3}.

\section{Beam Asymmetry Measurements}
Understanding the mechanisms of meson photoproduction is critical for disentangling the $J^{PC}$ quantum numbers of the observed states in the exotic hybrid mesons search. Theoretical models predict that the beam asymmetry $\Sigma$ is sensitive to the relative contributions from vector $1^{-}$ ($\rho^{0}/\omega$) and axial-vector $1^{+}$ ($b_{1}^{0}/h_{1}$) exchanges in $\pi^{0}$ and $\eta$ photoproduction~\cite{ref.4}.\\
The exclusive reactions $\gamma p \rightarrow p \pi^{0}$ and $\gamma p \rightarrow p \eta$ with $\pi^{0}/\eta\rightarrow \gamma\gamma$, are studied. The yields for the PERP and PARA orientations are given by
\begin{align}
Y_{\parallel/\perp} \propto N_{\parallel/\perp}[\sigma_{0}A(\phi)(1 \mp P \Sigma \cos 2\phi_{p})],
\label{eq.1}
\end{align}
where $\Sigma$ is the beam asymmetry, $N_{\parallel/\perp}$ is the flux of photons in two orthogonal orientations, $\sigma_{0}$ is the unpolarized cross section, $A(\phi)$ is an arbitrary function for the $\phi$-dependent detector acceptance and efficiency, $P$ is magnitude of the beam polarization and $\phi_{p}$ is the azimuthal angle of the production plane defined by the final-state proton.\\
The orthogonality of the PARA and PERP polarization configurations cancels out the $\phi$-dependent instrumental acceptance using
\begin{align}
\frac{Y_{\perp}-F_{R}Y_{\parallel}}{Y_{\perp}+F_{R}Y_{\parallel}} = \frac{(P_{\perp}+P_{\parallel})\Sigma \cos 2\phi_{p}}{2+(P_{\perp}-P_{\parallel})\Sigma \cos 2\phi_{p}},
\label{eq.2}
\end{align}
where $F_{R} = N_{\perp}/N_{\parallel}$ is the ratio of the integrated photon flux between PERP($N_{\perp}$) and PARA ($N_{\parallel}$). Figure~\ref{fig.3} shows the yield asymmetry as a function of $\phi_{p}$, which is fit using the functional form in Eq.(~\ref{eq.2}), where the beam asymmetry $\Sigma$ is the only free parameter.

\begin{figure}[h]
    \centering
    \includegraphics[width=16pc]{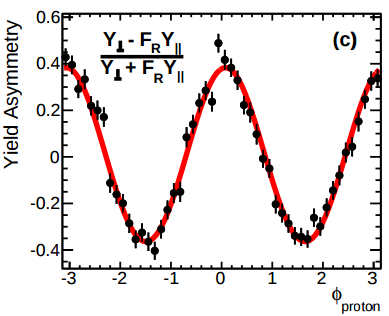}
    \caption{\label{fig.2}The yield asymmetry, fit with Eq.(\ref{eq.2}) to extract $\Sigma$.}
\end{figure}

The beam asymmetry is determined in bins of momentum transfer ($-t$) for $\pi^{0}$ and $\eta$ photoproduction, and the results are shown in Fig.~\ref{fig.3}.

\begin{figure}[h]
    \centering
    \begin{minipage}{16pc}
        \includegraphics[width=16pc]{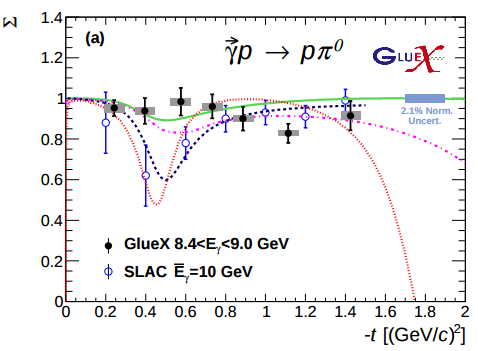}
    \end{minipage}\hspace{3pc}%
    \begin{minipage}{16pc}
        \includegraphics[width=16pc]{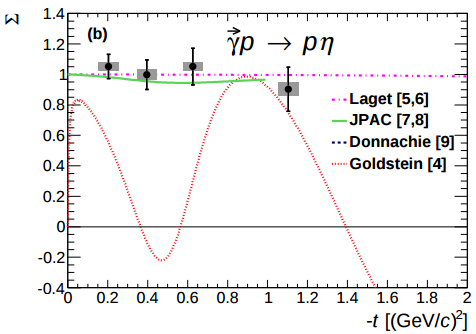}
    \end{minipage}
    \caption{\label{fig.3} Beam asymmetry $\Sigma$ for (a) $\gamma p \rightarrow p \pi^{0}$ and (b) $\gamma p \rightarrow p \eta$ (black filled circles). Uncorrelated systematic errors are indicated by the height of gray bars and combined statistical and systematic uncertainties are given by the black error bars. The previous SLAC results at $\overline{{E}}_{\gamma}$ = 10 $GeV$ (blue open circles) are also shown along with various Regge theory calculations (see ref.~\cite{ref.4} and references therein).}
\end{figure}

The theoretical models predict a dip near $-t$ = 0.5 $(GeV/c)^{2}$, due to contributions from axial-vector meson exchange matching previous $\pi^{0}$ measurements at $\overline{E}_{\gamma}$ = 10 $GeV$ from the Stanford Linear Accelerator Center (SLAC). This dip is not observed in the GlueX data, which strongly suggests the dominance of vector meson exchange at this energy. The $\eta$ beam asymmetry measurements are the first above 3 $GeV$, and are consistent with unity over the measured -t range.

\section{Charmonium Photoproduction Near Threshold}
\begin{figure}[h]
    \centering
    \begin{minipage}{16pc}
        \includegraphics[width=18pc]{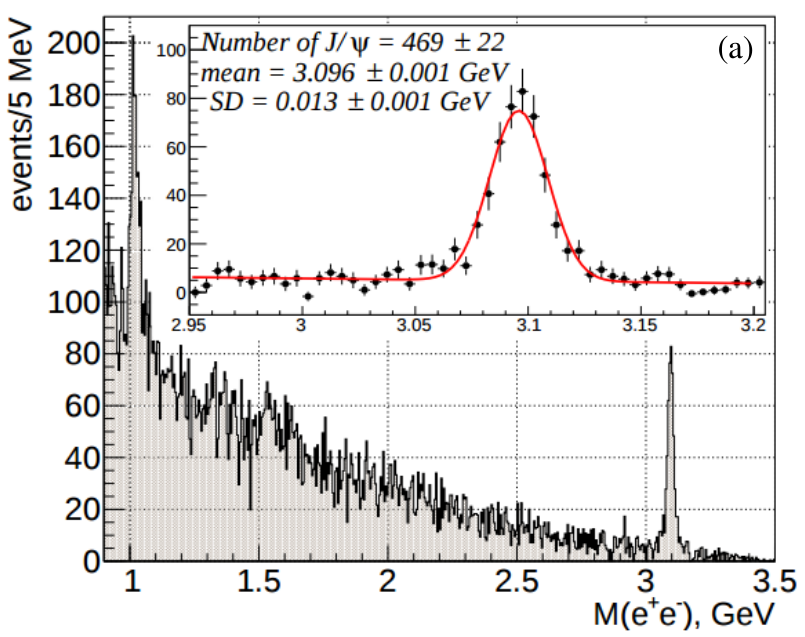}
    \end{minipage}\hspace{3pc}%
    \begin{minipage}{16pc}
        \includegraphics[width=18pc]{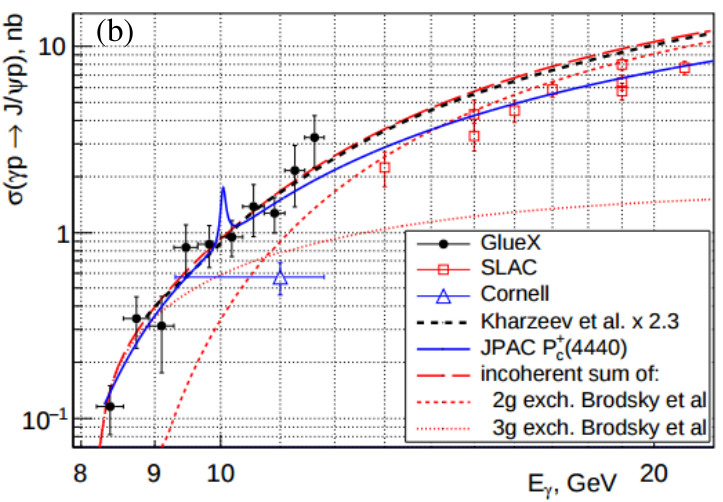}
    \end{minipage}
    \caption{\label{fig.4} (a) Electron-positron invariant mass spectrum from the data. The insert shows the $J/\psi$ region fitted with a linear polynomial plus a Gaussian (fit parameters shown). (b) GlueX results for the $J/\psi$ total cross section vs beam energy, together with the Cornell and SLAC data, the theoretical predictions, and the JPAC model (details found in~\cite{ref.5}). All curves are fitted/scaled to the GlueX data only. For our data the quadratic sums of statistical and systematic errors are shown; the overall normalization uncertainty is 27\%.}
\end{figure}

The study of near-threshold $J/\psi$ photoproduction can provide information on the gluonic structure of the nucleus, for example through measuring the contribution of the leading order (two gluon exchange) or higher twist (three gluon exchange) processes in the production mechanism~\cite{ref.5}. This process can also be used to search for the pentaquark candidates reported by the LHCb experiment in the $J/\psi p$ channel of the $\Lambda^{0}_{b} \rightarrow J/\psi p K^{-}$ decay~\cite{ref.5}.\\
The exclusive reaction $\gamma p \rightarrow p e^{+}e^{-}$ was selected, which includes the narrow $\phi$ and $J/\psi$ peaks, and a continuum dominated by the Bethe-Heitler (BH) process. Figure~\ref{fig.4} (a) shows the invariant mass spectrum of $e^{+}e^{-}$ data after the event selection. We normalize the $e^{+}e^{-}$ total cross section to that of BH in the invariant mass range $1.20$ - $2.50$ GeV, thus minimizing uncertainties from factors like luminosity and common detector efficiencies.

The measured total cross section in bins of beam energy is shown in Figure~\ref{fig.4} (b), compared to other measurements from photoproduction experiments and to the theoretical models. We find that our data do not favor either pure two- or three-hard-gluon exchange separately, and a combination of the two processes is required to fit the data adequately.
The narrow LHCb states, $P^{+}_{c}(4312)$, $P^{+}_{c}(4440)$, and $P^{+}_{c}(4457)$, produced in the s-channel would appear as structures at $E_{\gamma}$ = $9.44$, $10.04$ and $10.12$ GeV in the cross-section, but the results seen in Figure~\ref{fig.4} (b) show no evidence for such structures. We obtain model-dependent upper limit on the branching fraction at 90\% confidence level of 4.6\%, 2.3\%, and 3.8\% for $P^{+}_{c}(4312)$, $P^{+}_{c}(4440)$, and $P^{+}_{c}(4457)$ states, respectively.

\section{Hybrid Meson Photoproduction Search}
\begin{figure}[h]
    \centering
    \includegraphics[width=35pc]{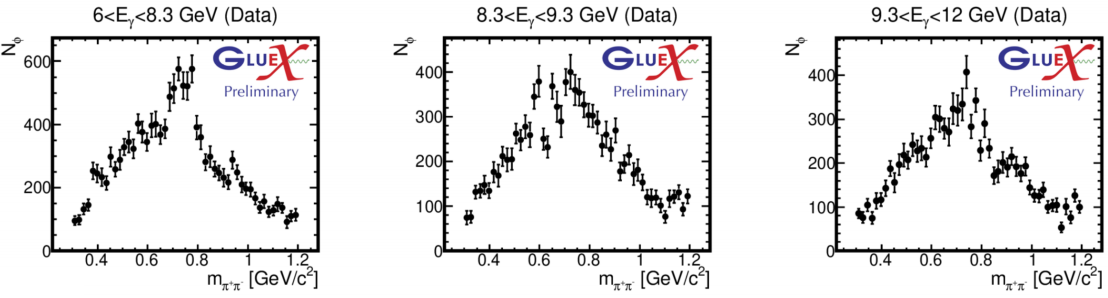}
    \caption{\label{fig.7}$\phi(1020)$ yield versus $\pi^{+}\pi^{-}$ invariant mass in 3 different bins of beam energy ($E_{\gamma}$).}
\end{figure}

\begin{figure}[h]
    \centering
    \includegraphics[width=35pc]{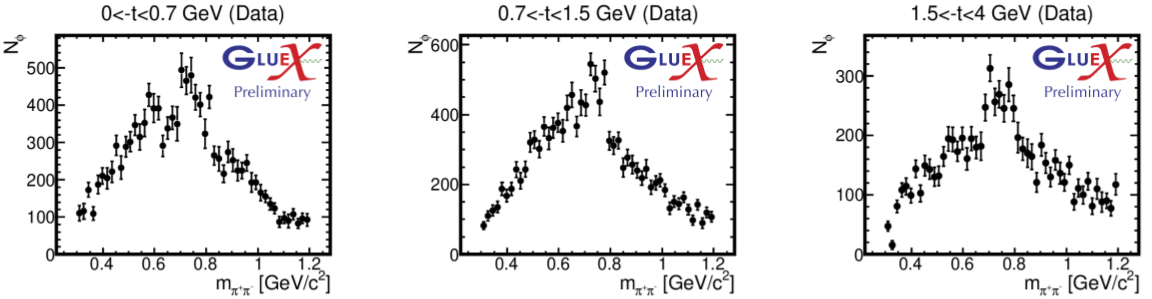}
    \caption{\label{fig.8}$\phi(1020)$ yield versus $\pi^{+}\pi^{-}$ invariant mass in 3 different bins of momentum transfer ($-t$).}
\end{figure}

One of the potential candidates for the $1^{--}$ hybrid mesons is the $Y(2175)$, a possible strangeonium counterpart of the $Y(4260)$ in the charmonium sector, which has been already observed in positron-electron experiments~\cite{ref.6}~\cite{ref.7}. The GlueX experiment offers a new opportunity to search for this state for the first time in photoproduction. Since the $Y(2175)$ is seen in $\phi(1020)f_{0}(980)$ and $\phi(1020)\pi^{+}\pi^{-}$ states, we study the exclusive reaction $\gamma p \rightarrow p \pi^{+}\pi^{-}K^{+}K^{-}$. In order to remove the background underneath the $\phi(1020)$ in the $K^{+}K^{-}$ invariant mass, we fit a $\phi(1020)$ signal plus background as a function of the $\pi^{+}\pi^{-}$ invariant mass and this way extract the $\pi^{+}\pi^{-}$ invariant mass-dependent $\phi(1020)$ yields at different beam energy (Figure~\ref{fig.7}) and momentum transfer bins (Figure~\ref{fig.8}). The contribution from $\rho(770)$ is significant, and an enhancement near the nominal $f_{0}(980)$ mass is seen, but not significant to claim the evidence. A similar method will be applied to extract the $\phi(1020)\pi^{+}\pi^{-}$ invariant mass-dependent $\phi(1020)$ yields in different bins, to search for the presence of the $Y(2175)$ state.

\section{Summary}
The first part of the GlueX program is to map the conventional meson spectrum, and understand the production mechanisms. Production of the pseudoscalars $\pi^{0}$ and $\eta$ is dominated by vector meson exchange. The same exchanges are expected to dominate in the production of $1^{-+}$ exotic hybrid mesons. An opportunistic study of charmonium production at threshold lead to upper limits of several percent on the branching fraction of the LHCb $P^{+}_{c}$ pentaquarks. A DIRC (Detection of Internally Reflected Cherenkov light) detector is currently installed for hadron identification, which will open a broader program to search for hybrid mesons with strangeness.

\end{document}